\documentstyle[12pt]{article}
\def\be{\begin{equation}}
\def\ee{\end{equation}}
\def\l{\label}

\def\F{{\cal F}}

\def\S{{\cal S}}

\def\T{{\cal T}}
\def\V{{\cal V}}
\def\Z{{{}}}
\def\H{{\cal H}}

\def\U{{\cal U}}
\def\W{{\cal W}}
\def\Q{{\cal Q}}
\def\V{{\cal V}}

\def\s{{\tt s}}
\def\t{{\tt t}}

\def\NPB#1#2#3{{\it Nucl.\ Phys.}\/ {\bf B#1} (19#2) #3}

\def\PRD#1#2#3{{\it Phys.\ Rev.}\/ {\bf D#1} (19#2) #3}
\def\PRL#1#2#3{{\it Phys.\ Rev.\ Lett.}\/ {\bf #1} (19#2) #3}

\begin{document}
\begin{titlepage} 

\rightline{\tt hep-th/9801033}
\rightline{UFIFT-HEP-98-1}
\rightline{DFPD97/TH/50}

\vspace{1.3cm}

\begin{center} 

{\Large \bf Quantum Transformations}

\vspace{1.cm}

 {\large Alon E. Faraggi$^{1}$ $\,$and$\,$ Marco Matone$^{2}$\\}
\vspace{.2in}
 {\it $^{1}$ Institute for Fundamental Theory, Department of Physics, \\
        University of Florida, Gainesville, FL 32611,
        USA\\
e-mail: faraggi@phys.ufl.edu\\}
\vspace{.025in}
{\it $^{2}$ Department of Physics ``G. Galilei'' -- Istituto 
                Nazionale di Fisica Nucleare\\
        University of Padova, Via Marzolo, 8 -- 35131 Padova, Italy\\
   e-mail: matone@padova.infn.it\\}

\end{center}

\vspace{0.8cm}

\centerline{\large Abstract}

\vspace{0.2cm}

\noindent
We show that the stationary quantum Hamilton--Jacobi equation of
non--relativistic 1D systems, underlying Bohmian mechanics, takes the
classical form with $\partial_q$ replaced by $\partial_{\hat q}$ where 
$d\hat q={dq\over \sqrt{1-\beta^2}}$. The $\beta^2$ term
essentially coincides with the quantum potential that, like $V-E$,
turns out to be proportional to a curvature arising in projective geometry.
In agreement with the recently formulated equivalence principle,
these ``quantum transformations'' indicate that
the classical and quantum potentials deform space geometry.

\end{titlepage}\newpage
\setcounter{footnote}{0} 
\renewcommand{\thefootnote}{\arabic{footnote}}

One of the main aspects of contemporary theoretical research concerns
the quantization of gravity. Despite many efforts and results, such as those of
superstring theory, the understanding of the problem is still incomplete.
While Einstein's general relativity, based on a simple principle, describes
gravity in a purely geometrical framework, foundations of quantum mechanics
rely on a set of axioms which apparently seem unrelated to any geometrical
principle. It is then natural to think that the difficulties which arise
in considering quantization of gravity merit a better understanding
of the possible relationship between the foundations of general relativity
and quantum mechanics.

Recently we proposed that quantum mechanics may in fact
arise from an equivalence principle
\cite{1}\cite{2}. While the original formulation considered the case of
non--relativistic one--dimensional stationary systems with Hamiltonian of the
form $H=p^2/2m+V(q)$, which is also the case we consider in the present Letter,
it will be shown in \cite{FMQM}\cite{BFM} that the principle actually implies
the higher dimensional time--dependent Schr\"odinger equation.

In this Letter we show that the Quantum Stationary Hamilton--Jacobi
Equation (QSHJE),
that we derived from the equivalence principle \cite{1}\cite{2},
maps to the classical form under ``quantum transformations" whose structure
is strictly related to the quantum potential. This indicates that the
classical and quantum potentials
deform space geometry. We will also show that both the quantum
potential and $V-E$ are proportional to curvatures arising in projective 
geometry. These aspects, together with the investigation
of $p$--$q$ duality, related to the properties of the Legendre
transformation, constitute the main results of the present Letter.

The solution $\S_0$ of the QSHJE
derived in \cite{1}\cite{2} is the quantum version of the
Hamiltonian characteristic function (also called reduced action).
In this respect the theory is consistently defined in terms of trajectories
\cite{Floyd1}\cite{2}\cite{FMQM}. Although reminiscent of Bohmian mechanics 
\cite{Bohm}\cite{Holland}, the formulation we consider has some
differences which will be further considered in \cite{FMQM}.
In particular, as noticed also by  Floyd \cite{Floyd1}, 
while in Bohm theory one identifies $\psi=Re^{{i\over \hbar}\S}$
with the Schr\"odinger wave function, one can see that 
in the 1D stationary case
the natural identification is $\psi=R(A e^{{i\over\hbar}\S_0}+B
e^{-{i\over\hbar}\S_0})$. While in Bohm theory
the state described by a real
wave function corresponds to $\S_0=0$, this is never the case in the
approach we consider. Furthermore, we note that the Schwarzian derivative
$\{\S_0,q\}$ is not defined for $\S_0=cnst$. As a consequence, while in
Bohm theory the states described by a real wave function unavoidably have a
vanishing conjugate momentum, this is never the case
in the proposed formulation.
While in Bohmian mechanics there is the issue of recovering the classical limit
for states with real wave function, e.g. for the harmonic oscillator in which
$\S_0=0$ \cite{Holland}, this limit is rather natural 
in the formulation that we consider
\cite{FMQM}. Another aspect is that there isn't any wave
guide in the proposed approach. Furthermore, a basic fact is that the conjugate
momentum $p=\partial_q\S_0$ is a real quantity even in classically
forbidden regions.
In \cite{FMQM} we will see that also the quantized energy spectra and their
structure are a direct consequence of the equivalence principle.

Let us consider two
1D stationary non--relativistic systems
with Hamilton's characteristic
functions $\S_0(q)$ and $\S_0^v(q^v)$. Setting
\be
\S_0^v(q^v)=\S_0(q),
\l{is1}\ee
induces the ``$v$--transformations''
\be
q\longrightarrow q^v=v(q),
\l{6xxx}\ee
where $v=\S_0^{{v}^{\;-1}}\circ \S_0$,
with $\S_0^{{v}^{\;-1}}$ denoting the inverse of $\S_0^v$. 

Recently, the following problem has been considered in \cite{1,2}

\vspace{.5cm}

\noindent
{\it Given an arbitrary system with reduced action $\S_0(q)$, 
find the coordinate transformation $q\longrightarrow  q^{v_0}=v_0(q)$,
such that the new reduced action $\S_0^{v_0}$, defined by
\be
\S_0^{v_0} (q^{v_0}) =\S_0(q),
\l{universalstate}\ee
corresponds to the free system with vanishing energy.}

\vspace{.5cm}

In the following we will use the notation
$q^0\equiv q^{v_0}$, $\S_0^{0}\equiv \S_0^{v_0}$.
We also set $\W(q)\equiv V(q)-E$,
and denote the state $\W=0$ by\footnote{By $\W$ states we will mean for short
the physical systems corresponding to a potential $V$ and energy $E$.}
\be
\W^0(q^0)\equiv 0. 
\l{def3}\ee
Observe that the structure of the states described by $\S_0^{0}$ and $\S_0$
determines the ``trivializing coordinate" $q^{0}$ to be
\be
q\longrightarrow q^{0} =\S_0^{{0}^{\;-1}}\circ \S_0(q),
\l{9thebasicidea}\ee

Let us denote by $\H$ the space of all possible ${\W}$'s. Since 
the approach extends to arbitrary physical systems, the
space $\H$ is a rather general one and may include cases in 
which $V(q)$ is a distribution.
In particular, even if the possible
potentials should be restricted to the ones physically realizable in nature, 
it is clear that the structure of this space cannot be defined a priori.
Rather, for a given potential $V(q)$, the possible values of $E$ are determined
by the properties of local homeomorphicity of the $v$--maps which are natural to
impose from the equivalence principle and that will be
discussed later. This principle,
suggested by the problem of finding the trivializing map (\ref{9thebasicidea}),
states that \cite{1}

\vspace{.5cm}

\noindent
{\it For each pair $\W^a,\W^b\in\H$, there is a $v$--transformation such that}
\be
\W^a(q)\longrightarrow {\W^a}^v (q^v)=\W^b(q^v).
\l{equivalence}\ee

\vspace{.5cm}

\noindent
Note that this implies that there always exists
the trivializing coordinate 
$q^{0}$ for which $\W\longrightarrow \W^0$. 
Let us consider the properties that the $v$--maps
should have in order that
the equivalence principle be satisfied.
First of all note that
$v$--maps should be continuous:
since both $q$ and $q^v$ take values
continuously on ${\bf R}$,
it is clear that full equivalence between the two systems requires that the
$v$--maps should be continuous. This is the general situation. However,
depending on the structure of the potential, it may happen that the physical
system is confined to an interval of the real line. This corresponds to a
degenerate case. In particular, in studying the structure of the conjugate
momentum $p=\partial_q\S_0$, the case of the infinitely deep well is
conveniently studied as a limiting procedure \cite{Floyd1}\cite{FMQM}.
The equivalence principle is still
satisfied with the trivializing map restricted to the finite interval
delimited by the turning points \cite{FMQM}.

Note that the equivalence principle implies that the transformation
(\ref{6xxx}) should exist for any couple of physical systems. This provides the
pseudogroup property (see below). In particular, one has to impose that the
$v$--transformations be locally invertible. However, in discussing the
properties of these maps, such as continuity, one should consider the extended
real line $\hat{\bf R}={\bf R}\cup\{\infty\}$. Actually, since there are no
reasons to restrict to global one--to--one self--maps of ${\bf R}$, the issue
of continuity of the $v$--maps forces us to consider $\hat{\bf R}$. This avoids
considering the fictitious discontinuity arising at the points $\pm \infty$, a
property related to the structure of the real line and not to the intrinsic
properties of the $v$--maps. Compactifying the real line allows us to select
and discard the transformations which are intrinsically discontinuous.
Therefore, the $v$--maps should be local homeomorphisms of $\hat{\bf R}$ into
itself. In \cite{FMQM} we will see that this property also follows from
the structure of the QSHJE.

To better understand the above aspect, it is useful to map the extended real
line to the unit circle by means of a Cayley transformation and then
consider the case of the trivializing map. While
$z=(q-i)/(-iq+1)$ spans $S^1$ once, $w=(q^0-i)/(-iq^0+1)$ runs continuously
around the unit circle.\footnote{This property implies
the quantization of the energy spectra without making use of the axiomatic
interpretation of the wave function \cite{FMQM}.}
Since the Cayley transformation is a global univalent
transformation, we have
that the $v$--maps induce local self--homeomorphisms of the unit circle.
An interesting property of the $v$--maps is that associated to any
physical state there is an integer number associated to the order of the
covering of the trivializing map \cite{2}.

We note that since
local homeomorphisms are closed under composition, it follows that
local homeomorphicity of any $v$--map also follows from 
local homeomorphicity of the trivializing map. A similar aspect
is called pseudogroup property. In this respect it is worth noting
that this is the property of holomorphic functions which one uses 
for defining a complex analytic structure: this implies that
the composition of two complex analytic local homeomorphisms is again a
complex analytic local homeomorphism (see for example \cite{Gunning}
and references therein).

In \cite{1} it has been shown that the equivalence
principle implies the quantum analogue
of the Hamilton--Jacobi equation which in turns implies the Schr\"odinger
equation. Subsequently, it has been shown in \cite{2} that this is the
unique possible solution. Let us shortly review the structure of the 
derivation in \cite{1}\cite{2}. First of all one observes the basic fact that 
the equivalence principle cannot be consistently implemented in
classical mechanics. This can be summarized in the following steps

\begin{itemize} 
\item[{\bf 2)}]{Consider the Classical Stationary Hamilton--Jacobi Equation
(CSHJE) $(\partial_q\S_0^{cl} )^2=-2m\W$.
Given another system with reduced action $\S_0^{cl\, v}$, denote by $q^v$ the
new space coordinate and set $q^v=v(q)$, with $v$ determined by $\S_0^{cl\, v}
(q^v)=\S_0^{cl}(q)$, that is $v=\S_0^{{cl\, v}^{\;-1}}\circ \S_0^{cl}$;}
\item[{\bf 2)}]{compare the CSHJE for the system with reduced action $\S_0^{cl\,
v}$, that is $(\partial_{q^v}\S_0^{cl\, v}(q^v))^2=-2m\W^v(q^v)$, with
$(\partial_q\S_0^{cl}(q))^2=-2m\W(q)$, and use $\S_0^{cl\, v}(q^v)=\S_0^{cl}(q)$
so that $\W^v(q^v)=(\partial_qq^v)^{-2}\W(q)$. Hence, consistency implies
that in classical mechanics $\W$ belongs to $\Q$, the space of functions
transforming as quadratic differentials under $v$--maps;}
\item[{\bf 3)}]{the fact that in classical mechanics one has $\W \in\Q$, implies
that the state $\W^0$ is a fixed point in $\H$, i.e. under a coordinate
transformation $\W^0(q^0)\longrightarrow (\partial_{q^0}q^v)^{-2}\W^0(q^0)=0$.}
\end{itemize}

It is therefore clear that in order to implement the equivalence
principle the CSHJE should be modified. The most general form
would be
\be
{1\over 2m}\left({\partial {\S}_0(q)\over \partial q}\right)^2+\W(q)+Q(q)=0.
\l{aa10bbbb}\ee
Since classical mechanics exists, it is clear that the above equation must
reduce to the CSHJE in a suitable limit. That is in some limit
we must have
\be
Q\longrightarrow 0.
\l{classicoqezero}\ee
Since the equivalence principle
implies that $\W\notin\Q$, it is clear that 
classical mechanics is the covariance breaking phase with
$Q$ having the role of covariantizing term.

The properties of $\W+Q$ under the $v$--transformations 
are determined by the transformed equation $\left({\partial_{q^v} 
{\S}_0^v(q^v)}\right)^2/2m +\W^v(q^v)+Q^v(q^v)=0$, that by (\ref{is1})
and (\ref{aa10bbbb}) yields
\be
\W^v(q^v)+Q^v(q^v)=\left({\partial_q q^v}\right)^{-2}\left(\W(q)+Q(q)\right),
\l{yyyxxaa10bbbb}\ee
that is 
\be
(\W+Q)\in\Q.
\l{fiof}\ee

Let us recall how $Q$ is determined by the equivalence 
principle \cite{1}\cite{2}.
We have seen that if $\W$ transforms as a 
quadratic differential, then $\W^0$
would be a fixed point in the $\H$ space. It follows that
$\W\notin \Q$ so that by (\ref{fiof})
$Q\notin \Q$.
Therefore
\be
\W^v(q^v)=\left({\partial_{q^a} q^v}\right)^{-2}\W^a(q^a)+\Z(q^a;q^v),
\l{azzoyyyxxaa10bbbb}\ee
and by (\ref{fiof})
\be
Q^v(q^v)=\left({\partial_{q^a}  q^v}\right)^{-2}Q^a(q^a)-\Z(q^a;q^v).
\l{azzo2yyyxxaa10bbbb}\ee
For $\W^a(q^a)=\W^0(q^0)$
Eq.(\ref{azzoyyyxxaa10bbbb}) gives
\be
\W^v(q^v)=\Z(q^0;q^v).
\l{ddazzoyyyxxaa10bbbb}\ee
This means that all the states correspond
to the inhomogeneous part of the transformation of the state $\W^0$
induced by some coordinate transformation.

Let $a,b$ and $c$ denote arbitrary $v$--transformations. Comparing
\be
\W^b(q^b)=\left({\partial_{q^b}q^a}\right)^{2}\W^a(q^a)+\Z(q^a;q^b)=
\Z(q^0;q^b),
\l{ganzate}\ee
with the same formula 
with $q^a$ and $q^b$ interchanged we have
$\Z(q^b;q^a)=-(\partial_{q^a}q^b)^{2}\Z(q^a;q^b)$,
in particular
$\Z(q;q)=0$.
More generally, comparing
$$
\W^b(q^b)= \left({\partial_{q^b}  q^c}\right)^{2} \W^c(q^c)+\Z(q^c;q^b)=
 \left({\partial_{q^b}  q^c}\right)^{2}\left[
\left({\partial_{q^c}  q^a}\right)^{2}\W^a(q^a)+\Z(q^a;q^c)\right]+\Z(q^c;q^b)=
$$
\be
\left({\partial_{q^b}q^a}\right)^{2}\W^a(q^a)+
\left({\partial_{q^b}q^c}\right)^{2}
\Z(q^a;q^c)+\Z(q^c;q^b),
\l{DemetrioStratosZ}\ee
with (\ref{ganzate}), we obtain the basic relation \cite{2}
\be
\Z(q^a;q^c)=\left({\partial_{q^c}  q^b}\right)^{2}\Z(q^a;q^b)-
\left({\partial_{q^c}  q^b}\right)^{2}\Z(q^c;q^b),
\l{cociclo3}\ee
which extends to higher dimensions \cite{FMQM}\cite{BFM}.
This relation, which is a cocycle condition and directly follows from
the equivalence principle, actually implies \cite{2} 
\be
\Z(q^a;q^b)= -{\beta^2\over 4m} \{q^a,q^b\},
\l{equazione3}\ee
where $\beta$ is a dimensional constant and
\be
\{h(x),x\}={{h{'''}(x)}\over {h{'}(x)}}-
{3\over 2}\left({h{''}(x)\over {h{'}(x)}}\right)^2=
(\ln h'(x))''-{1\over 2}[(\ln h'(x))']^2,
\l{Schwarzian}\ee
denotes the Schwarzian derivative.
Since the inhomogeneous term in the transformation of $\W$
must disappear in the classical limit, we have by
(\ref{equazione3}) that the classical phase corresponds
to the $\beta\longrightarrow 0$ limit.
By (\ref{ddazzoyyyxxaa10bbbb}) and (\ref{equazione3}) it follows that
$\W$ itself is a Schwarzian derivative
\be
\W^v(q^v)=-{\beta^2\over 4m}\{q^0,q^v\},
\l{bastanza}\ee
with $q^0$ determined by 
the fact that the $\beta\longrightarrow 0$ limit corresponds to
the classical phase. One obtains \cite{1}\cite{2}
\be
Q={\beta^2\over 4m}\{\S_0,q\}.
\l{sothat2}\ee
Eq.(\ref{aa10bbbb}) and 
the identity\footnote{This identity admits a higher dimensional extension 
\cite{BFM}.}
\begin{equation}
\left({\partial_q{\S}_0}\right)^2=
{\beta^2\over 2}  \{e^{{2i\over\beta}{\S}_0},q\}-{\beta^2\over 2} 
\{{\S}_0,q\},
\label{expoid}\end{equation}
imply that Eq.(\ref{sothat2}) is equivalent to
\be
\W=-{\beta^2\over 4m}\{e^{{2i\over\beta}{\S}_0},q\}.
\l{sothat}\ee
By (\ref{aa10bbbb}) and (\ref{sothat2}) it follows
that the equation for $\S_0$ we were looking for is \cite{1}\cite{2}
\be
{1\over 2m}\left({\partial {\S}_0(q)\over \partial q}\right)^2+\W(q)
+{\beta^2\over 4m}\{\S_0,q\}=0,
\l{aa10bbbxxxb}\ee
which is equivalent to (\ref{sothat}). It follows that
\begin{equation}
e^{{{2i}\over\beta}{\S}_0}={A\psi^D +B\psi\over C\psi^D+D\psi}, 
\label{dfgtp}\end{equation}
$AD-BC\ne 0$, with $\psi^D$ and $\psi$
linearly independent solutions of the stationary Schr\"odinger equation
\begin{equation}
\left(-{\beta^2\over 2m}
{\partial^2\over \partial q^2}+V(q)\right)\psi=E\psi.
\label{yz1xxxx4}\end{equation}
Thus, for the ``covariantizing parameter'' we have
\be
\beta=\hbar,
\l{Planck}\ee
where $\hbar=h/2\pi$ and $h$ is the Planck constant. We note that
the QSHJE (\ref{aa10bbbxxxb})
has been already considered in literature
\cite{Messiah}\cite{Floyd1}\cite{Holland}.

In Ref.\cite{1} the function $\T_0(p)$, defined as 
the Legendre transformation of the reduced action, has been introduced
\be
\S_0(q)=pq-\T_0(p).
\l{y1}\ee
While $\S_0(q)$ is the momentum generating function, its Legendre 
dual $\T_0(p)$ is the coordinate generating function
\be
p={\partial \S_0\over \partial q} ,\qquad 
q={\partial \T_0\over \partial p}.
\l{y2}\ee

The second derivative of (\ref{y1}) with respect
to $\s=\S_0(q)$ yields the ``canonical equation''
\begin{equation}
\left(\partial^2_{\tt s} +{\U}({\tt s})\right)q\sqrt p
=0=\left(\partial^2_{\tt s}+{\U}({\tt s})\right)\sqrt p,
\label{10}\end{equation}
with the ``canonical potential" being
\be
{\U}({\tt s})=\{q\sqrt p/\sqrt p,{\tt s}\}/2=\{q,{\tt s}\}/2.
\l{edt5}\ee
Observe that the choice of the coordinates $q$ and $q^v$, which of course
does not imply any loss of generality as both $q$ and $q^v$ play the role 
of independent coordinate in their own system, allows
us to look at the reduced action as a scalar function.
In particular, since
$\S^v_0(q^v)=\S_0(q)$, we see that the transformations (\ref{6xxx}) leave the
Legendre transformation of $\T_0$ (\ref{y1}) unchanged.
Consequently, from
$\partial_{q^v}\S_0^v(q^v)=
\left( \partial_q q^v\right)^{-1} \partial_{q}\S_0(q)$, we have
$p\longrightarrow p_v=
\left( \partial_q q^v\right)^{-1} p$.
However, while the Legendre transformation of $\T_0$ is, by definition, 
invariant under
$v$--transformations, this is not the case for the canonical potential
$\U$. Nevertheless, there is an important exception as
under the $GL(2,{\bf C})$ transformations
\begin{equation}
q^v ={(Aq+B)/ (Cq+D)}\qquad \longrightarrow \qquad p_v = \rho^{-1}(Cq+D)^2p,
\label{4zxz}\end{equation}
$\rho=AD-BC\ne 0$, we have that the M\"obius symmetry of the
Schwarzian derivative implies
\be
{\U}({\tt s})=\{(Aq+B)/(Cq+D),\s\}/2=\U(\s).
\l{Mobiusssyy}\ee
Therefore we can speak of $GL(2,{\bf C})$--symmetry of the canonical equation.

Involutivity of the Legendre transformation and the duality
$$
{\S}_0\longleftrightarrow{\T}_0,\quad ~~~{q}\longleftrightarrow p,
$$
imply another $GL(2,C)$--symmetry, 
with the dual version of Eq.(\ref{10}) being \be
\left(\partial^2_t+{\V}({\t})\right)p\sqrt q
=0=\left(\partial^2_{\tt t}+{\V}({\tt t})\right)\sqrt q,
\l{abbdual}\ee
where
\be
{\V}({\tt t})=\{p\sqrt q/\sqrt q,{\tt t}\}/2=\{p,{\tt t}\}/2,
\l{fi}\ee
with $\t=\T_0(p)$.
We note that for $p=\gamma/q$ 
the solutions of 
(\ref{10}) and (\ref{abbdual}) coincide. Therefore we have the self--dual states
\be
\S_0=\gamma\ln \gamma_q q,\qquad {\T}_0=\gamma\ln \gamma_p p,
\l{selfduals}\ee
where the three constants satisfy 
\be
\gamma_p\gamma_q\gamma=e.
\l{tregamme}\ee
It will be shown in \cite{FMQM} that this equation
is connected to fundamental constants.
Note that
\begin{equation}
\S_0+\T_0=pq=\gamma,\qquad\U({\tt s})=-{1/4\gamma^2}=\V({\tt t}).
\label{s0t0}\end{equation}
We observe that the canonical equation (\ref{10}) and its dual (\ref{abbdual}) 
correspond to two equivalent descriptions of physical systems
that for the self--dual states overlap. Later we will consider another derivation
of the self--dual states (\ref{selfduals}).

Remarkably, the QSHJE (\ref{aa10bbbxxxb}) can be also seen as modification by a
``conformal factor'' of the CSHJE. In particular, using the identity
$$
\{q,\S_0\}=-(\partial_q 
\S_0)^{-2}\{\S_0,q\},
$$
we have that the canonical potential determines the conformal rescaling
\cite{1}\cite{2}
\begin{equation}
{1\over 2m}\left({\partial\S_0\over \partial q}\right)^2
\left[1-\hbar^2 \U(\S_0)\right]+V(q)-E=0.
\label{hsxdgyij}\end{equation}
This shows the basic role of the purely quantum mechanical self--dual states
(\ref{selfduals}). Actually, observe
that for the state $\W^0(q^0)$, Eq.(\ref{hsxdgyij}) has the form
\begin{equation}
{1\over 2m}\left({\partial\S_0^0\over \partial q^0}\right)^2
\left[1-\hbar^2 \U(\S_0^0)\right]=0.
\label{hsxdgyij0}\end{equation}
Setting
\be
\tilde q^0= {Aq^0+B\over Cq^0+D},
\l{mobiusdellacoordinata}\ee
the solution of (\ref{hsxdgyij0}) has the form
\be
\S_0^0={\hbar\over 2i}\ln \tilde q^0,
\l{lacuisoluzioneesebborea}\ee
that is
\be
1-\hbar^2\U({\hbar\over 2i}\ln\tilde q^0)=0.
\l{poidjq}\ee
In the case in which $\gamma_q=(A/D)^{\pm 1}$, $B=C=0$, the solution
(\ref{lacuisoluzioneesebborea}) corresponds to the two self--dual
states defined by (\ref{selfduals}) with 
\be
\gamma=\gamma_{sd}\equiv\pm{\hbar/2i}.
\l{piuomeno}\ee

The solution $\S_0^0={\hbar\over 2i}\ln\tilde q^0$ solves the problem of
finding the trivializing coordinate for which $\W(q)\longrightarrow \W^0(q^0)$.
Actually, by (\ref{9thebasicidea}) and (\ref{lacuisoluzioneesebborea}) we have
\be
\S_0^0(q^0)={\hbar\over 2i}\ln\left({Aq^0+B\over Cq^0+D}\right)=\S_0(q),
\l{belloabbasatanza}\ee
that is
\be
q\longrightarrow q^{0} =
{De^{{2i\over\hbar}\S_0(q)}-B\over -Ce^{{2i\over \hbar}\S_0(q)}+A}.
\l{9thebasicideabbb}\ee
We remark that related interesting issues have been recently considered 
in \cite{Periwal}.

Similarly to the case of general relativity in which
the equivalence principle leads to the deformation of the geometry, 
also in quantum mechanics one should investigate whether
the equivalence principle implies a space deformation. 
In this context, the structure  of the QSHJE (\ref{hsxdgyij}) suggests
considering an underlying geometrical
structure. Actually, Eq.(\ref{hsxdgyij}) naturally leads to a 
coordinate transformation depending on the quantum potential. The key point
is that (\ref{hsxdgyij}) can be written in the form
\begin{equation}
{1\over 2m}\left({\partial\S_0\over \partial \hat q}\right)^2+V(q)-E=0,
\label{hsxdgyijccc}\end{equation}
where
\be
\left({\partial q\over \partial \hat q}\right)^2=\left[1-\hbar^2 \U(\S_0)\right],
\l{5}\ee
or equivalently (we omit the solution with the minus sign) 
\be
d\hat q= {dq\over \sqrt{1-\beta^2(q)}},
\l{6}\ee
with
\be
\beta^2(q)= {\hbar^2}\U(\S_0)={\hbar^2\over 2}\{q,\S_0\}.
\label{beta}\ee
Integrating (\ref{5}) yields
\be
\hat q=\int^q{dx\over \sqrt{1-\beta^2(x)}}.
\l{quantumcoordinate}\ee
We observe that the nature of the
coordinate transformation is purely quantum mechanical; 
in particular
\be
\lim_{\atop\hbar\longrightarrow 0} \hat q=q.
\l{333}\ee

Equation (\ref{quantumcoordinate}) indicates that in considering the
differential structure one should take into account the effect of the
quantum potential on the space geometry.
In this context, the deformation of the CSHJE
amounts to replacing the standard derivative
with respect to the classical coordinate $q$ with the derivative
with respect to the deformed quantum coordinate $\hat q$. In other words,
the transition from the classical to the quantum regime
amounts to a reconsideration of the underlying geometry
which is modified by the quantum potential itself. 

A property of the quantum transformation (\ref{6}) is that
it allows to put the QSHJE in the
classical form. Namely, setting
\be
\hat \W(\hat q)=\W(q(\hat q)),
\l{1998a}\ee
\be
\hat \S_0(\hat q)=\S_0(q(\hat q)),
\l{1998b}\ee
it follows that Eq.(\ref{hsxdgyij}), equivalent to Eq.(\ref{hsxdgyij0}), can
be written in the form
\begin{equation}
{1\over 2m}\left({\partial\hat \S_0(\hat q)\over \partial \hat q}\right)^2+
\hat\W(\hat q)=0.
\label{bbhsxdgyijccc}\end{equation}
This can be seen as the opposite of the problem,
considered by Schiller and Rosen \cite{SchillerRosen},
of determining the wave function representation
for classical mechanics (see also \cite{Holland}).

In the standard formulation of the
quantum analogue of the Hamilton--Jacobi equation \cite{LL}, 
one considers a couple
of equations which arise by setting $\psi=Re^{i\S_0/\hbar}$, so that
for the state $\W^0(q^0)$
one chooses $\S_0^0=cnst$ and $R=Aq^0+B$. Note that
setting $\psi=Re^{i\S_0/\hbar}$ is suggested by the interpretation
of $|\psi|^2=R^2$ as probability density. On the other hand, it is easy to see
that any solution has the form
\be
\psi={1\over \sqrt{\S_0'}}\left(A e^{-{i\over \hbar}\S_0}+
Be^{{i\over \hbar}\S_0}\right).
\label{popca}\ee
However, while on the one hand
it is not possible to define the Legendre transformation of a constant, so that
the $\S_0$--$\T_0$ duality would be lost and $\{\S_0^0,q^0\}$ cannot be defined, 
on the other hand, we have that the
overlooked solution $\S_0^0= {\hbar\over 2i}\ln\tilde q^0$ for the state
$\W^0$ still gives the same solution $\psi=Aq^0+B$ of the Schr\"odinger
equation $-{\hbar^2\over 2m}\partial_q^2\psi=0$. In this context we
observe that the non--linear relation between $\S_0$ and the wave function,
which can be also written in the form $\S_0(q)={\hbar\over 2i}
\ln(A\int^q\psi^{-2}+B)/(C\int^q\psi^{-2}+D)$, is related to an
incomplete equivalence between the Schr\"odinger equation and the 
QSHJE (\ref{aa10bbbxxxb}). Another interesting example of
inequivalence between the Schr\"odinger equation and Eq.(\ref{aa10bbbxxxb})
is provided in \cite{Floyd1}\cite{Floyd2} where it has been shown that for
bound states the QSHJE (\ref{aa10bbbxxxb}) describes microstates
not detected in the Schr\"odinger representation. This aspect
will be considered in great detail in \cite{FMQM}. 

The fact that the QSHJE admits the 
classical representation (\ref{bbhsxdgyijccc}) suggests that
classically forbidden regions correspond to critical regions
for the quantum coordinate. Actually, writing 
Eq.(\ref{quantumcoordinate}) in the equivalent form ($\s=\S_0(q)$)
\be
\hat q=\int^qdx{\partial_x\S_0\over \sqrt{-2m\W}}=
\int^{\S_0(q)}{d\s\over \sqrt{-2m\W}},
\l{quantumcoordinatebbb}\ee
we see that the integrand is purely 
imaginary in the classically forbidden regions $\W>0$.
Furthermore, since according to (\ref{hsxdgyij0})
the conformal factor for the state $\W^0$ vanishes, it follows
by (\ref{quantumcoordinate}) that the quantum coordinate for the free
particle state with vanishing energy is divergent.
To better understand the role of the state
$\W^0$ in (\ref{quantumcoordinatebbb})
it is useful to first rederive the self--dual states 
(\ref{selfduals}) by another approach.

The $\S_0$--$\T_0$ duality implies that a given physical system may be 
described either by the $\S_0$--picture or by the $\T_0$--picture.
On general grounds, it is clear that naturally selected states
are the ones corresponding to the degenerate case in which 
the $\S_0$ and $\T_0$ pictures overlap. In order to find this common 
subspace
we consider the interchange of the $\S_0$ and $\T_0$ pictures given by
\be
q\longrightarrow \tilde q=\alpha p \qquad p\longrightarrow 
\tilde p=\beta q.
\l{sd1}\ee
This implies that
\be
{\partial \tilde \T_0 \over \partial \tilde p}=\alpha {\partial 
\S_0\over \partial q}, \qquad 
{\partial \tilde \S_0 \over \partial \tilde q}=\beta {\partial 
\T_0\over \partial p},
\l{oiqhx}\ee
which is equivalent to
\be
{\partial \tilde \T_0 \over \partial q}=\alpha\beta {\partial 
\S_0\over \partial q}, \qquad 
{\partial \tilde \S_0 \over \partial p}=\alpha\beta {\partial 
\T_0\over \partial p},
\l{zzoiqhx}\ee
that is
\be
\tilde \S_0(\tilde q) = \alpha \beta \T_0 (p) + cnst, \qquad 
\tilde \T_0(\tilde p) = \alpha \beta \S_0 (q) + cnst.
\l{sd111}\ee
Furthermore, since we require that (\ref{sd1}) be of order two,
we have up to an additive constant
\be
{\tilde {\tilde \S_0}} = \S_0, \qquad  {\tilde {\tilde \T_0}} = \T_0, 
\l{sd112}\ee
so that
\be
(\alpha \beta)^2 = 1.
\l{aridaiieerizzummolo}\ee
We observe that $\tilde \S_0(\tilde q)$ and $\tilde \T_0(\tilde p)$ are 
basically the Legendre transformation of $\S_0 (q)$ and $\T_0(p)$ respectively.
The distinguished states are precisely those which are left invariant by 
(\ref{sd1}) and (\ref{sd111}), that is
\be
\tilde \S_0(\tilde q) = \S_0 (q) + cnst.
\l{sd114}\ee

Let us now introduce the Legendre 
transformation of the Hamilton principal function $\S$
\begin{equation}
{\S}=p{{\partial {\T}\over \partial p}}-{\T},\qquad
{\T}=q{{\partial {\S}\over \partial q}}-{\S},
\label{lt00tempo}\end{equation}
\begin{equation}
p={{\partial {\S}}\over \partial q},\qquad \qquad q=
{{\partial {\T}}\over\partial p}.
\label{pqtempo}\ee
Observe that for stationary states 
\be
\S(q,t)=\S_0(q)-Et,\qquad \T(p,t)=\T_0(p)+Et.
\l{timeind}\ee
Let us consider the differentials
\be
d\S={\partial \S\over \partial q}dq+{\partial \S\over \partial t}dt=
pdq+{\partial \S\over \partial t}dt,
\l{diffs}\ee
\be
d\T={\partial \T\over \partial p}dp+{\partial \T\over \partial t}dt=
qdp+{\partial \T\over \partial t}dt,
\l{difft}\ee
so that
\be
d\S=d(pq-\T)=pdq+qdp-qdp-{\partial \T\over \partial t}dt,
\l{implicanoa}\ee
that is
\be
{\partial \S\over \partial t}=-{\partial \T\over \partial t}.
\l{zummolo}\ee
This equation connects the $\S$ and $\T$ pictures through
the time evolution. By (\ref{sd111}) 
(\ref{aridaiieerizzummolo}) (\ref{sd114}) and (\ref{timeind})
we have that
the distinguished states correspond to
\be
\S=\pm \T+cnst.
\l{sd6}\ee
As (\ref{sd6}) should 
be stable under time evolution, the relation (\ref{zummolo})
fixes the sign ambiguity and sets 
\be
\alpha \beta = -1.
\l{cazzarola}\ee 
Therefore, the distinguished states correspond to
\be
\S=-\T+cnst.
\l{sd7}\ee
Since $\S=pq-\T$,
we have
\be
pq=\gamma,
\l{sd11}\ee
where $\gamma$ is a constant. Therefore, the distinguished states are
precisely the self--dual states (\ref{selfduals}).

We have seen that the self--dual states with 
$\gamma=\gamma_{sd}\equiv\pm{\hbar/2i}$
correspond to the state $\W^0$.
The fact that it corresponds to two of the distinguished
states connecting the $\S_0$ and $\T_0$ pictures, indicates
that $\W^0$ corresponds to a critical point for the coordinate
transformation. In this context the observed divergence for $\hat q$ 
corresponding to this state is not a surprise.

We note that due to the M\"obius symmetry of the Schwarzian 
derivative, one has
\be
\W=-{\hbar^2\over 4m}\{e^{{2i\over\hbar}{\S}_0},q\}=
-{\hbar^2\over 4m}\left\{{Ae^{{2i\over\hbar}{\S}_0}+B\over
Ce^{{2i\over\hbar}{\S}_0}+D},q\right\}.
\l{sothatxcv}\ee
This means that to find $\S_0$ we need to fix three integration constants.
Let us set
\be
w={\psi^D\over \psi},
\l{deltapar}\ee
with $\psi^D$ and $\psi$ two linearly independent real solutions of the 
Schr\"odinger
equation (note that these solutions always exist).
Reality condition for $\S_0$ restricts (\ref{dfgtp}) to
\be
e^{{2i\over \hbar}\S_0}=e^{i\alpha} {w+i\bar\ell\over w-i\ell},
\l{KdT3}\ee
with $\alpha$ and $\ell$ real and complex integration constants respectively.
Note that ${\rm Re}\,\ell\ne 0$.
The solution (\ref{KdT3}), derived in
\cite{2}, is an equivalent form of the one previously derived in \cite{Floyd1}.

It follows by (\ref{expoid}) that what is invariant is
not $\left({\partial_{q}\S_0}\right)^2$ but rather
$\left({\partial_q{\S}_0}\right)^2+{\hbar^2}\{{\S}_0,q\}/2$.
This shows that the ``kinetic term'' ${1\over 2m}\left({\partial_{q}\S_0}
\right)^2$ and the quantum correction $Q={\hbar^2\over 4m} 
\{{\S}_0,q\}$ mix under a change of initial conditions. A property
of this mixing is that this disappears  
in the classical regime only, where $Q\longrightarrow 0$.
We note that the role of the quantum correction $Q$ is somehow 
reminiscent of the relativistic rest energy, as it 
is an intrinsic property of the particle. 

The above investigations, and the equivalence principle in particular,
indicate that quantum mechanics is strictly connected to
geometrical properties of space. It it then natural to investigate
the existence of possible geometrical structures underlying
the QSHJE. In order to do this we 
use a result obtained by Flanders \cite{Flanders} who showed that
the Schwarzian derivative can be interpreted as an invariant (curvature)
of an equivalence problem for curves in ${\bf P}^1$.

Let us introduce a frame for ${\bf P}^1$. This consists
of a pair ${\bf x},{\bf y}$ of points in affine space ${\bf A}^2$
such that $[{\bf x},{\bf y}]=1$, where $[{\bf x},{\bf y}]=x_1y_2-x_2y_1$
is the area function. Considering
the moving frame $s\longrightarrow \{{\bf x}(s),{\bf y}(s)\}$
and differentiating $[{\bf x},{\bf y}]=1$ yields the structure equations
\be
{\bf x}'=a{\bf x}+b{\bf y}, \qquad {\bf y}'=c{\bf x}-a{\bf y},
\l{equazionistrtturali}\ee
where $a,b$, and $c$ depend on $s$. Given a map $\phi=\phi(s)$ from a domain
to ${\bf P}^1$, one can choose a moving frame ${\bf x}(s),{\bf y}(s)$ in such
way that $\phi(s)$ is represented by ${\bf x}(s)$. Observe that this map
can be seen as a curve in ${\bf P}^1$. Two mappings $\phi$ and $\psi$ are said 
to be equivalent if $\psi=\pi \circ\phi$ with $\pi$ a projective transformation
on ${\bf P}^1$. 

Flanders considered two extreme situations.
Let $b(s)=0$, for all $s$. In this case $\phi$ is constant.
Actually, taking the derivative of
$\lambda{\bf x}$, for some $\lambda(s)\ne 0$, by
(\ref{equazionistrtturali}) we have $(\lambda {\bf x})'=
(\lambda'+a\lambda){\bf x}$.  Choosing
$\lambda\propto exp [{-\int^s_{s_0}dta(t)}]\ne 0$,
we have $(\lambda {\bf x})'=0$, so that
$\lambda {\bf x}$ is a constant representative of $\phi$. 

The other case is for $b$ never vanishing. There are
only two inequivalent situations. The first one is when $b$ is either
complex or positive.
It turns out that it is always possible to choose
the following ``natural moving frame" for $\phi$ \cite{Flanders}
\be
{\bf x}'={\bf y},\qquad {\bf y}'=-k{\bf x}.
\l{primo}\ee
In the other case in which $b$ is real and negative, the natural moving frame
for $\phi$ is
\be
{\bf x}'=-{\bf y},\qquad {\bf y}'=k{\bf x}.
\l{secondo}\ee

A characterizing property of the
natural moving frame is determined up to a sign and $k$
is an invariant. Thus, for example, suppose that for a given $\phi$
there is, besides (\ref{primo}), the natural moving frame
${\bf x}'_1={\bf y}_1$, ${\bf y}'_1=-k_1{\bf x}_1$. Since both
${\bf x}$ and ${\bf x}_1$ are representatives of $\phi$, we have
${\bf x}=\lambda{\bf x}_1$, so that ${\bf y}={\bf x}'=
\lambda'{\bf x}_1+\lambda {\bf y}_1$ and $1=[{\bf x},{\bf y}]=\lambda^2$.  
Therefore, ${\bf x}_1=\pm {\bf x}$, ${\bf y}_1=\pm {\bf y}$
and $k_1=k$ \cite{Flanders}.

Let us now review the derivation of Flanders formula for $k$.
Consider $s\longrightarrow {\bf z}(s)$ to be an affine representative of
$\phi$ and let ${\bf x}(s),{\bf y}(s)$ be a natural frame.
Then ${\bf z}=\lambda{\bf x}$ where $\lambda(s)$ is never vanishing.
Now note that, since ${\bf z}'=\lambda'{\bf x}+\lambda{\bf y}$, we have
that $\lambda$ can be written in terms of the area function
$[{\bf z},{\bf z}']=\lambda^2$.
Computing the relevant area functions, one can check that
that $k$ has the following expression
\be
2k={[{\bf z},{\bf z}'''] +3[{\bf z}',{\bf z}'']\over [{\bf z},{\bf z}']}-
{3\over 2}\left({[{\bf z},{\bf z}'']\over [{\bf z},{\bf z}']}\right)^2.
\l{k}\ee

Given a function $z(s)$, this can be seen as the non--homogeneous coordinate
of a point in ${\bf P}^1$. Therefore, we can associate to $z$ the map
$\phi$ defined by $s\longrightarrow (1,z(s))={\bf z}(s)$. In this case
we have $[{\bf z},{\bf z}']=z'$,  $[{\bf z},{\bf z}'']=z''$, 
$[{\bf z},{\bf z}''']=z'''$,  
$[{\bf z}',{\bf z}'']=0$, and the curvature becomes \cite{Flanders}
\be
k={1\over 2}\{z,s\}.
\l{rimarchevole}\ee

Let us now consider a state $\W$. We have
\be
\W= -{\hbar^2\over 4m} \{e^{{2i\over\hbar}{\S}_0},q\}=
-{\hbar^2\over 2m}k_\W.
\l{sothatW}\ee
Similarly, for the quantum potential
\be
Q={\hbar^2\over 4m} \{{\S}_0,q\}= {\hbar^2\over 2m}k_Q,
\l{sothatQ}\ee
where $k_\W$ is the curvature associated to the map 
\be
q\longrightarrow (1,e^{{2i\over\hbar}{\S}_0(q)}),
\l{splendido}\ee
while the curvature $k_Q$ is associated to the map
\be
q\longrightarrow (1,{\S}_0(q)).
\l{splendidoDemetrioStratosArea}\ee
The function defining the map (\ref{splendido}) corresponds
to the one defining the trivializing map (\ref{9thebasicideabbb}).

The identification of $-2m\W/\hbar^2$ with the curvature $k_\W$
allows us to write
the Schr\"odinger equation in the geometrical form
\begin{equation}
\left({\partial^2\over \partial q^2}+k_\W \right)\psi=0.
\label{polbuabuicina}\end{equation}
Furthermore,
the identity (\ref{expoid}) can be now seen as difference of curvatures
\begin{equation}
\left({\partial_q{\S}_0}\right)^2={\hbar^2}k_\W-{\hbar^2}k_Q,
\label{expoidQW}\end{equation}
and the QSHJE (\ref{aa10bbbxxxb}) can be written in the form
\be
{1\over 2m}\left({\partial {\S}_0(q)\over\partial q}\right)^2+\W(q)
+{\hbar^2\over 2m}k_Q=0.
\l{aa10bbbxxxbcurv}\ee

Let us now consider the meaning of the natural moving frame in the framework
of the QSHJE. First observe that the structure equations imply that
\be
{\bf x}''=-k {\bf x}.
\l{movingff}\ee

In the case of $k=k_\W$, this equation is the Schr\"odinger equation, so that
\be
{\bf x}=(\psi^D,\psi),\qquad{\bf y}=(\psi^{D'},\psi'),
\l{xypsidpsi}\ee
and the frame condition is nothing else but the statement that the
Wronskian $W$ of the Schr\"odinger equation is a constant
\be
[{\bf x},{\bf y}]=\psi^D\psi'-\psi\psi^{D'}=W=1.
\l{wroschiano}\ee
Hence, the Schr\"odinger equation determines
the natural moving frame associated to the curve in ${\bf P}^1$
given by the representative (\ref{splendido}) with $-2m\W/\hbar^2$
denoting the invariant associated to the map.
In other words, the Schr\"odinger problem corresponds
to finding the natural moving frame such that
$-2m\W/\hbar^2$ be the invariant curvature.

In the case $k=k_Q$, Eq.(\ref{movingff}) becomes
\be
\left({\hbar^2\over 2m}{\partial^2\over \partial q^2}+Q\right)\phi=0,
\l{IY2}\ee
so that if $\phi^D$ and $\phi$ are solutions of (\ref{IY2}), then
\be
\S_0= {A\phi^D+B\phi\over C\phi^D+D\phi}.
\l{IY3}\ee
Note that the solutions of (\ref{IY3}) are related
to the solutions of the Schr\"odinger equation by 
\be
\gamma_\psi(\psi^D/\psi)=e^{{2i\over\hbar}\gamma_\phi(\phi^D/\phi)},
\l{IY4}\ee
where $\gamma_\psi$ and $\gamma_\phi$ denote two M\"obius transformations.

Let us conclude this Letter with a few remarks.
As we noticed above, with respect to the standard solution for the free quantum
state with vanishing energy, $\S_0^0=cnst$, which is the same as for classical
mechanics, one has that what is vanishing is not $\partial_{q^0}\S_0^0$ but
$2(\partial_{q^0} {\S}_0^0(q^0))^2+\hbar^2
\{\S_0^0,q^0\}=0$, so that $\S_0^0={\hbar\over 2i}\ln
\tilde q^0$ and there is a non zero curvature term associated to
the free particle with vanishing energy. 

In the conventional approach to quantum mechanics the discretized spectra and
its structure arise from the properties imposed on the wave function.
For example, for the harmonic oscillator one requires that
the wave function vanishes at infinity: a direct consequence of the axiomatic
interpretation of the wave function as probability amplitude.
An outcome of \cite{FMQM} is that the quantized energy spectra and their
structure are a direct consequence of the equivalence principle.
Therefore, two basic aspects of quantum mechanics, such as the tunnel effect and
energy quantization (and its structure),
strictly related to the wave function interpretation,
arise in our approach as a consequence of the equivalence principle.

While this principle has been formulated for the 1D case, it
actually implies the time dependent Schr\"odinger equation
in $D+1$ dimensions. The point is that
this is the unique possibility if one requests that 
the deformed Hamilton--Jacobi equation reduces to the classical one in the 
$\hbar\longrightarrow 0$ limit and reproduces $D$ copies of the
one--dimensional quantum Hamilton--Jacobi equation in the case in which
$V(q_1,\ldots,q_D,t)=\sum_{k=1}^DV_k(q_k)$. In doing this,
one uses the higher dimensional generalizations of
the cocycle condition (\ref{cociclo3})
and of the identity (\ref{expoid}) \cite{FMQM}\cite{BFM}.

It is worth stressing that in the higher dimensional case the resulting
quantum Hamilton--Jacobi equation reproduces the Bohmian one but with the 
some additional important conditions \cite{FMQM} such as the exclusion
of the solution $\S_0=\sum_{k=1}^DA_kq_k+B$.
This aspect follows from a detailed analysis which includes
the study of both the $E\longrightarrow 0$ and $\hbar\longrightarrow 0$
limits \cite{FMQM}. This fact is strictly related to the
existence of the Schwarzian derivative and
of the Legendre transformation of $\S_0$,
which in turn is related to the issue of $p$--$q$ duality \cite{FMQM}.

We observe that our investigation is related to the approach in \cite{FM},
further
developed by Carroll in \cite{Carroll}, where it has been shown that the space
coordinate is proportional to the Legendre transformation
of the prepotential $\F$, defined
by $\psi^D=\partial_\psi\F$, with respect to the square of the wave function.

Finally, we note that very recently related issues have been considered
in \cite{Anandan}.

\vspace{.333cm}

\noindent
It is a pleasure to thank G. Bertoldi, G. Bonelli, E.R. Floyd and M. Tonin for 
interesting discussions.
Work supported in part by DOE Grant No.\ DE--FG--0586ER40272 (AEF)
and by the European Commission TMR programme ERBFMRX--CT96--0045 (MM).


\newpage

\begin{thebibliography}{99}

\bibitem{1} A.E. Faraggi and M. Matone, hep-th/9705108. 
\bibitem{2} A.E. Faraggi and M. Matone, hep-th/9711028, to appear
in {\it Phys. Lett.} {\bf B}.
\bibitem{FMQM} A.E. Faraggi and M. Matone, {\it The Equivalence
Postulate of Quantum Mechanics}, in preparation.
\bibitem{BFM} G. Bertoldi, A.E. Faraggi and M. Matone, in preparation.
\bibitem{Floyd1} E.R. Floyd,  
\PRD{\bf 25}{82}{1547};
{\bf D26} (1982) 1339; {\bf D29} (1984) 1842; {\bf D34} (1986) 3246;
{\it Phys. Lett.} {\bf 214A} (1996) 259.
\bibitem{Bohm} D. Bohm, Phys. Rev. {\bf 85} (1952) 166; ibidem 180.
\bibitem{Holland} P.R. Holland, {\it The Quantum Theory of Motion},
Cambridge Univ. Press (Cambridge, 1993).
\bibitem{Gunning} R.C. Gunning, {\it Lectures on Riemann Surfaces},
Princeton Univ. Press, (Princeton, 1966).
\bibitem{Messiah} A. Messiah, {\it Quantum Mechanics}, Vol.1, North--Holland
(Amsterdam, 1995).
\bibitem{Periwal} V. Periwal, \PRL{80}{98}{4366}, hep-th/9709200.
\bibitem{SchillerRosen} R. Schiller, {\it Phys. Rev.} 
{\bf 125} (1962) 1100; ibidem
1109.\\
A. Rosen, {\it Am. J. Phys.} {32} (1964) 597; {\it Found. Phys.}
 {\bf 16} (1986) 687.
\bibitem{LL} L.D. Landau and E.M. Lifschitz, {\it Quantum Mechanics}, 
Pergamon Press (Oxford, 1958).
\bibitem{Floyd2} E. Floyd, {\it Found. Phys. Lett.} {\bf 9} (1996) 489,
quant-ph/9707051.
\bibitem{Flanders} H. Flanders, {\it J. Diff. Geom.} {\bf 4} (1970) 515.
\bibitem{FM} A.E. Faraggi and M. Matone,  \PRL{78}{97}{163},
hep-th/9606063.
\bibitem{Carroll} R. Carroll, hep-th/9607219; hep-th/97010216;
                              hep-th/9702138; \NPB{502}{97}{561}, 
hep-th/9705229.
\bibitem{Anandan} J.S. Anandan, gr-qc/9712015.

\end{thebibliography}
\end{document}